\documentclass[fleqn,usenatbib]{mnras}
\usepackage{newtxtext,newtxmath}
\usepackage[T1]{fontenc}
\usepackage{units}
\usepackage{lineno}
\usepackage{color}
\usepackage{epsfig}
\usepackage{float}
\usepackage{bm}
\DeclareRobustCommand{\VAN}[3]{#2}
\let\VANthebibliography\thebibliography
\def\thebibliography{\DeclareRobustCommand{\VAN}[3]{##3}\VANthebibliography}
\usepackage{graphicx} 
\usepackage{amsmath}	 

\usepackage{amssymb}	 
\usepackage{amsfonts}
\usepackage{caption}
\usepackage{subfigure}

\title[Plate Collision Model]{Repeating Fast Radio Bursts with High Burst Rates by Plate Collisions in Neutron Star Crusts}

\author[Li et al.]{
Qiao-Chu Li,$^{1, 2}$
Yuan-Pei Yang,$^{3}$\thanks{E-mail: ypyang@ynu.edu.cn (YPY)}
F. Y. Wang, $^{1,2}$
Kun Xu$^{4,5}$ 
and Zi-Gao Dai$^{6,1}$\thanks{E-mail: daizg@ustc.edu.cn (ZGD)}
\\
$^{1}$School of Astronomy and Space Science, Nanjing University, Nanjing 210023, China\\
$^{2}$Key laboratory of Modern Astronomy and Astrophysics (Nanjing University), Ministry of Education, Nanjing 210023, China\\
$^{3}$South-Western Institute for Astronomy Research, Yunnan University, Kunming 650500, China \\
$^{4}$School of Astronomy and Space Sciences,  University of Chinese Academy of Sciences, Beijing, China\\
$^{5}$Key Laboratory of Optical Astronomy, National Astronomical Observatories, Chinese Academy of Sciences, Beijing, China\\
$^{6}$Department of Astronomy, School of Physical Sciences, University of Science and Technology of China, Hefei 230026, China
}

\date{Accepted XXX. Received YYY; in original form ZZZ}

\pubyear{2022}

\begin{document}
\label{firstpage}
\pagerange{\pageref{firstpage}--\pageref{lastpage}}
\maketitle

\begin{abstract}
Some repeating fast radio burst (FRB) sources show high burst rates, and the physical origin is still unknown.
Outstandingly, the first repeater FRB 121102 appears extremely high burst rate with the maximum value reaching $122\,\mathrm{h^{-1}}$ or even higher. 
In this work, we propose that the high burst rate of an FRB repeater may be due to plate collisions in the crust of young neutron stars (NSs).
In the crust of an NS, vortex lines are pinned to the lattice nuclei.
When the relative angular velocity between the superfluid neutrons and the NS lattices is nonzero, a pinned force will act on the vortex lines, which will cause the lattice displacement and the strain on the NS crust growing.
With the spin evolution, the crustal strain reaches a critical value, then the crust may crack into plates, and each of plates will collide with its adjacent ones. 
The Aflv\'en wave could be launched by the plate collisions and further produce FRBs.
In this scenario, the predicted burst rate can reach $\sim 770\,\mathrm{h}^{-1}$ for an NS with the magnetic field of $10^{13}\,\unit{G}$ and the spin period of $0.01\,\unit{s}$.
We further apply this model to FRB 121102, and predict the waiting time and energy distribution to be $P(t_{\mathrm{w}}) \propto t_{\text{w}}^{\alpha_{t_{\text{w}}}}$ with $\alpha_{t_{\text{w}}} \simeq -1.75$ and $N(E)\text{d}E \propto E^{\alpha_{E}}\text{d}E$ with $\alpha_{E} \simeq -1.67$, respectively. These properties are consistent with the observations of FRB 121102.
\end{abstract}

\begin{keywords}
(transients:) fast radio bursts -- (stars:) pulsars:general 
\end{keywords}

\section{Introduction}
Fast radio bursts (FRBs) are mysterious radio transients with millisecond durations and extremely high brightness temperatures \citep[e.g.,][]{cordes2019, petroff2019, xiaodi2021}. 
Since the first FRB was discovered in 2007 \citep{lorimer2007}, hundreds of FRBs have been detected \citep[e.g.,][]{frbcatnew2020, chimecat2021}. 
Remarkably, FRB 200428 \citep{bochenek2020, chime20200428} has been detected to be associated with an X-ray burs from Galactic magnetar SGR J1935+2154 \citep{mereghetti2020}, which indicates that at least a part of FRBs are originated from magnetars.
Recently, the repeating FRB 20200120E has been localized in a globular cluster of M81  with an old stellar environment, which means there is more than one origin of FRBs \citep{bhardwaj2021, kirsten2021}.
However, the radiation mechanism of FRBs is still unknown \citep[e.g.,][]{lyubarsky2014, dai2016, katz2016, murase2016, beloborodov2017, kumar2017, lyutikov2017, waxman2017, zhang2017, yang2018, metzger2019, wadiasingh2019, beloborodov2020, dai2020, geng2020, lu2020, margalit2020, wangweiyang2020, wu2020, xiao2020, yang2020, zhang2020mech, wangweihua2021, yang2021, yu2021}. 

In particular, some FRB sources can exhibit repeating behaviors \citep[e.g.,][]{spitler2016, chime180814}, while others can only be observed once even after long-term observations \citep[e.g.,][]{keane2016, bhandari2018, shannon2018, petroff2019FRB110214, qiuhao2019}.
The former are generally called repeating FRBs, and the latter are one-off FRBs.
According to the First CHIME/FRB Catalog \citep{chimecat2021}, there are some significant differences in the pulse width and bandwidth for one-off and repeating FRBs  \citep{widthbandwidth2021}.
It remains an open question whether 
one-off FRBs are one kind of repeaters with a very low burst rate \citep[e.g.,][]{connor2020}.

Notably, the first repeater, FRB 121102, is characterised by a possible $\sim 160$ days periodic activity \citep{rajwade2020, cruces2021}, and the high burst rate with a peak burst rate of about  $122\,\unit{h}^{-1}$ or even higher \citep{lidi21, jahns2022}.
Recently, there is another active FRB, FRB 20201124A, also having a high burst rate of about $45.8\,\unit{h}^{-1}$ \citep[e.g.,][]{nimmo2021, xu2021}. 
The burst rates of repeating FRB sources have been proposed to be due to the crust fracturing of an NS in some previous works \citep[e.g.,][]{yang2021, liqiaochu2021}.
In the simulation of the crust of young magnetars, the evolution of magnetic fields in the crust can cause crustal failures \citep{dehman2020}.
\cite{yang2021} suggested that the burst rate is related to the event rate of crust fracturing of a magnetar due to the shear stress by magnetic fields, and the bursts are mainly from the fragile regions near the field poles.
\cite{liqiaochu2021} studied the burst rate caused by starquakes due to the spin evolution of the neutron star (NS) in a binary model, with the interaction between the NS magnetosphere and the accreted material. 

For NSs with spin evolution, i.e., spin down or spin up, the shear stresses in the crust lattices would grow due to vortex lines pinned to lattice nuclei.
\cite{ruderman1, ruderman2, ruderman3} proposed that when these shear stresses are beyond the elastic yield strength in the NS crust, it would cause large-scale crust cracking events;
furthermore, for a rapidly spinning NS, this shear stresses will break and move the crust before vortex unpinning occurs.
However, whether the failure of the NS crust is plastic flow or brittle fracture has remained controversial so far \citep[e.g.,][]{blaes1989, thompson95, thompson1996, ruderman1998, jones2003, horowitz09, levin2012}.
A brittle crust means that there are fractures and displacements of the crust, and a plastic crust implies that it has plastic deformation without cracking.
\cite{thompson2000} pointed out that the mechanical failures of the NS crust consist of brittle fractures and plastic deformation. 
When the crust temperature is lower than $0.1T_{\text{m}}$ (where $T_{\text{m}}$ is the crystal melting temperature), the crust can change from plastic flow to brittle response \citep{ruderman2}. 
We show that this model is consistent with brittle fracture in Section \ref{sec2.1}.
On the other hand, some theories suggest that cracks and voids may not form in the NS crust \citep[e.g.,][]{jones2003, horowitz09, levin2012}, and the main reason is brittle fractures of the NS crust might heal quickly.
\cite{jones2003} argued that, under the huge isotropic pressure greater than the shear modulus of the NS, no long-lived void is formed after cracks. Most of the local stress energy may not be converted to kinetic energy, but transferred to neighboring regions. 
Thus only plastic deformation might occur.
In their molecular simulations, \cite{horowitz09} found that voids heals quickly under the extreme pressure in the crust.
And some global simulations of NS crusts are based on plastic flows in the entire crust \citep[e.g.,][]{kojima2020, gourgouliatos2021}.
We note that there are still some doubts about no long-lived void. 
First, when the NS crust cracks, the isotropic pressure may suddenly change and not be so strong that concentrated zones of strong shear are ubiquitous according to the study of \cite{thompson2017} and thus the lifetime of voids can be longer, and the crust will not heal quickly after cracking.
Second, some oscillation modes with a quantity of energy can make cracks be amplified, so that the lifetime of voids becomes longer \citep{mcDermott1988, lihuiquan2022}.
Third, for a very young NS or magnetar, the crust may crack more easily than an older NS \citep{suvorov2019}.

In this work, we propose that the plate collisions in the crust of a young NS, with a fast spin velocity, high glitch amplitude and strong magnetic field, can account for the high burst rate, waiting time distribution and energy distribution, which is consistent with the observations of FRB 121102. 
We analyze the physical process of the crust cracking due to the vortex lines pinned to the crust lattice nuclei in Section \ref{sec2.1}. 
We discuss the burst rate and the triggering mechanism of repeating FRBs in the scenario of vortex pinning-induced crust cracking in Section \ref{sec2.2}.
We calculate the distribution of the waiting time and energy caused by the plates cracking and collision in Section \ref{sec2.3}. 
Finally, we make some
discussion and conclusions in Section \ref{sec3}.

\section{Plate Collision Model}\label{sec2}
The crack of the crust of an NS could be caused by many reasons: 
the strong magnetic field in the NS \citep[e.g.,][]{thompson95, yang2021}, starquakes \citep[e.g.,][]{ruderman1972}, and the vortex pinning-induced crust cracking \citep{ruderman1}.  
When the stress in the crust of an NS reaches a critical value, the crust could crack.
In this work, we focus on the plate fragmentation and movement caused by the vortex pinning with the spin evolution of an NS.
The plates of the NS crust will collide with each other after the crust cracking, and further produce FRBs via the Alfv\'en wave \citep{kumar2020}.
In the following discussion, we will analyze the physical process of the plate collision. Then we will give an estimation of the burst rate, energy and duration of repeating FRBs triggered in this process.
Finally, we will study the distributions of waiting time and energy of a repeating FRB source.

\subsection{Vortex Pinning-induced Crust Cracking}\label{sec2.1}

Because of the extremely high density of NSs, their interiors are in superfluid states \citep[e.g.,][]{migdal1959, sedrakian2019}. 
The rotating superfluid forms vortex lines, each of which
carries a quantum of angular momentum \citep[e.g.,][]{epstein1988}.
In the crust of the NS with a relatively lower density, the superfluid material coexists with a crystalline lattice of the normal material.
Vortex lines are pinned to the lattice nuclei of the crust of the NSs \citep{anderson1975}.

The relative rotation velocity between the angular velocities of the superfluid neutrons, $\boldsymbol{\Omega_{\rm n}}$, and NS lattice, $\boldsymbol{\Omega}$, with superfluid vortex lines pinned can be represented by
\begin{align}
\boldsymbol{\omega} \equiv \boldsymbol{\Omega_{{\rm n}}} - \boldsymbol{\Omega}.
\end{align}
The vortex lines are pinned to the crust lattice nuclei. 
If the angular velocity of the superfluid neutrons and the crust of the NS are equal, there is no stress on the crust lattice nuclei within the pinning region of the crust. 
Otherwise, a force density would act on the vortex lines \citep{ruderman1},
\begin{align}
\boldsymbol{\mathcal{F}} = 2\boldsymbol{\omega} \times (\boldsymbol{\Omega_{{\rm n}}} \times \boldsymbol{r})\rho_{{\rm n}},
\end{align}
where $\boldsymbol{r}$ is the position of the vortex lines, and $\rho_{\rm n} \sim 2.8 \times 10^{14} \mathrm{~g} \mathrm{~cm}^{-3} $ is the density of the superfluid neutron. 

We define $\theta_{\text v}$ as the angle between the position of the vortex lines and the spin axis of the NS, and $s(\theta_{\text v})$ as the tangential displacement of the crust due to the vortex pinned force. For a small change of the angle $\text{d}\theta_{\text v}$, the tangential displacement of the crust is $\text{d} s$.
Therefore, the strain of the NS crust due to the vortex pinned force density can be estimated by \citep{ruderman1}
\begin{equation}
\varepsilon \sim \frac{\text{d} s}{R_{\text{NS}} \text{d} \theta_{\text v}}=\frac{\omega \Omega_{\rm n} \rho_{\rm n} R_{\text{NS}}^{2} }{12 \mu} \cos 2 \theta_{\text v},
\label{strain}
\end{equation}
where $R_{\text{NS}} = 10^6\,\unit{cm}$ is the radius of the NS.

The maximum stress that the NS crust can sustain before cracking can be represented by $\sigma_{\rm m} = \varepsilon_{\rm m} \mu$, where the breaking strain $\varepsilon_{\rm m}$ ranges from $10^{-5}$ to about $10^{-1}$ \citep{smoluchowski70, horowitz09}, and $\mu$ is the shear modulus of the NS crust of about $10^{30}\,\unit{g\,cm^{-1}\,s^{-2}}$ \citep{thompson95, douchin01, piro05}.
The crust cracks when $\varepsilon \sim \varepsilon_{\rm m}$, and the relative rotation velocity reaches a critical value of \citep{ruderman1},
\begin{align}
\omega_{B} &= \frac{12 \varepsilon_{\rm m } \mu}{\left|1-2 \sin ^{2} \theta_{\text v}\right|\rho_{\rm n} \Omega_{\rm n} R_{\text{NS}}^{2}} \notag \\
& \sim 3.4 \times 10^{-4}\,\mathrm{rad\,s}^{-1}\left(\frac{\varepsilon_{\rm m }}{10^{-5}}\right)
\left(\frac{P_{\text{NS}}}{ 10^{-2}\,\mathrm{s}}\right),
\end{align}
with the NS spin period $P_{\text{NS}}$.
According to Equation.(\ref{strain}), the size of a plate can be estimated by \citep{ruderman1}
\begin{equation}
R_{\rm P} \sim R_{\text{NS}}\left|\frac{\omega_{B}}{\Omega(t)-\Omega\left(t_{0}\right)}\right|^{1 / 2},
\end{equation}
where $\Omega(t)$ and $\Omega(t_0)$ are the angular velocities of the NS after the crust fracture, and with the organized vortex pinning, respectively.
Based on the observations of radio pulsars, the glitch amplitude $\Delta \Omega/\Omega$, with the difference between the angular velocity after and before the glitch in pulsars, is in the range from $10^{-12}$ to $10^{-4}$ \citep{espinoza11}.
Notably, there is a correlation between the pulsar  characteristic age and the glitch amplitude, i.e., a young pulsar tends to have a larger glitch amplitude \citep[e.g.,][]{wangn2000, espinoza11, basu21}.
Here, we estimate that $\left|\Omega(t)-\Omega\left(t_{0}\right)\right|/\Omega(t) \sim \Delta \Omega/\Omega$.
Therefore, $R_{\rm p}$ can be estimated by
\begin{equation}
R_{\rm P} \sim 7.4\times 10^3\,\unit{cm}\left(\frac{P_{\text{NS}}}{10^{-2}\,\unit{s}}\right)\left(\frac{\varepsilon_{\rm m }}{10^{-5}}\right)^{1/2}
\left(\frac{\Delta \Omega/\Omega}{10^{-2}}\right)^{-1/2}.
\label{rp}
\end{equation}

\begin{figure}
\centering
\includegraphics[width=\columnwidth, trim = 90 0 0 0, clip]{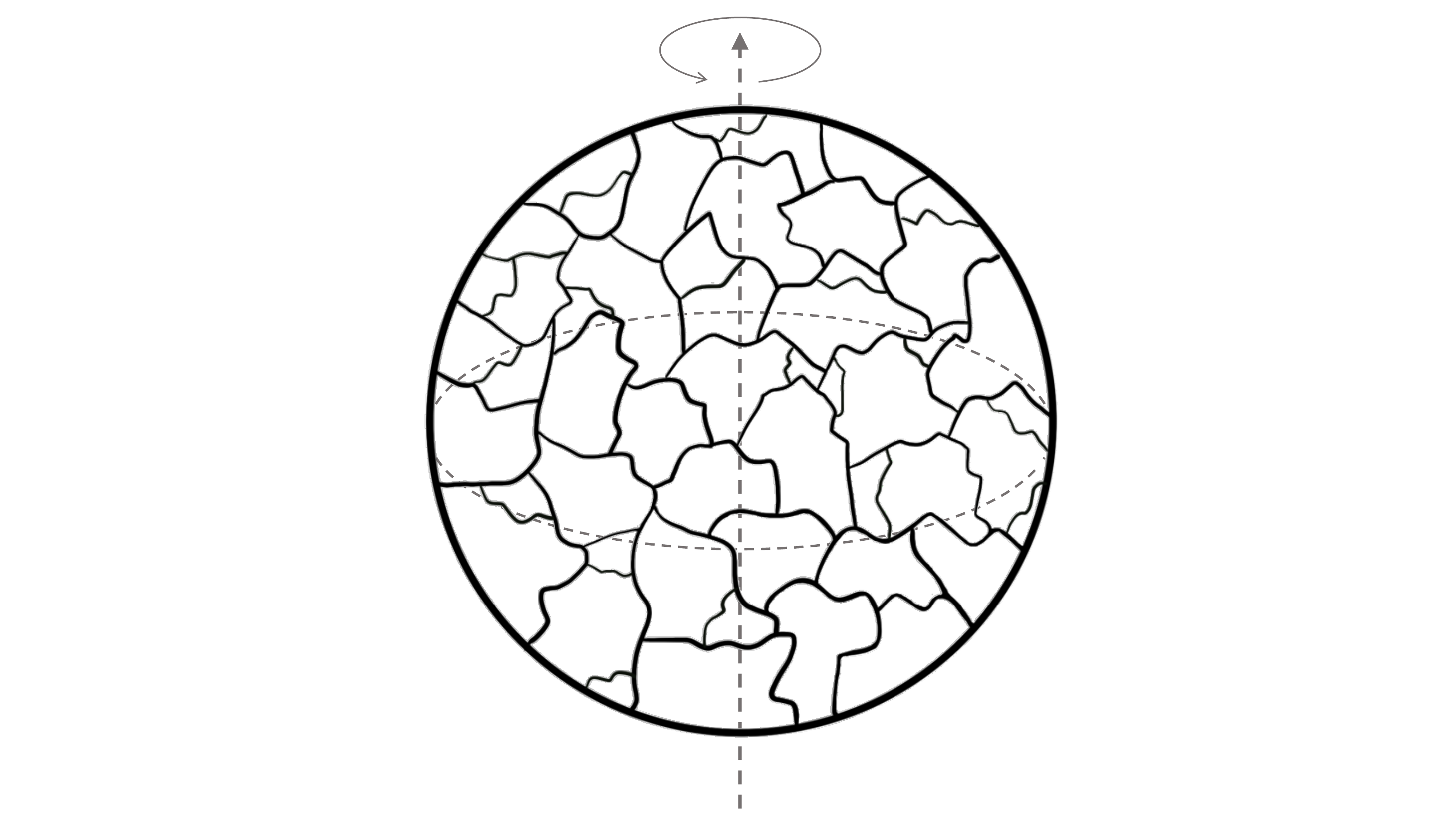}
\caption{Schematic illustration of the cracking crust of the NS in this work. 
The dotted line with an arrow represents the rotation axis of the NS.
When the stress in the crust of the NS reaches a critical value, the crust cracks into many plates with a size distribution.
They will move and collide with each other.}
\label{plate model}
\end{figure}

Compared with the nucleus outside the vortex core, the energy of the vortex line pinned by the crust nucleus is reduced.
When the force density of the relative rotation velocity between the superfluid neutrons and NS lattice is over the pinning force, then a vortex core contained nucleus can be pulled apart, therefore, there will be the vortex unpinning from the nucleus.
And the critical relative rotation velocity to make vortex lines unpinning from the lattice nuclei in the NS satisfies \citep{alpar1984, epstein1988, ruderman1}
\begin{equation}
\omega_{c}=\frac{\hbar R_{\rm N}^{2}}{\pi R_{\text{NS}} \sin \theta_{\text v}  m_{\rm n} b_{z}^{3}}
\sim \frac{8.2}{\sin \theta_{\text v}} \,\unit{rad\,s^{-1}},
\end{equation}
where $R_{N} = 7\times 10^{-13}\,\unit{cm}$ and $b_z = 2.2\times 10^{-12}\,\unit{cm}$ \citep{negele73} are the radius and the separation of the lattice nuclei at a density of $\sim 2.8 \times 10^{14} \mathrm{~g} \mathrm{~cm}^{-3}$, respectively.

\subsection{Repeating FRBs Triggered by Plate Collision of NS}\label{sec2.2}

After the NS crust cracking, the plates will collide with each other.
The collision rate can be estimated by the motion of plates.
The average tangential velocity of plates to the equator of the NS is \citep{ruderman1, ruderman3}
\begin{align}
v_{t} = & \frac{-\dot{\Omega} R_{\text{NS}} r_{\perp}}{2 \Omega\left(R_{\text{NS}}^{2}-r_{\perp}^{2}\right)^{1 / 2}} \simeq 5.0 \times 10^{-4}\,\unit{cm\,s^{-1}} \tan \theta \notag \\
& \times \left(\frac{\dot P_{\text{NS}}}{10^{-11}\,\unit{s\,s^{-1}}}\right)\left(\frac{P_{\text{NS}}}{10^{-2}\,\unit{s}}\right)^{-1},
\label{vt}
\end{align}
taking the change rate of angular velocity of the NS $|\dot{\Omega}|= 2 \pi P_{\text{NS}}^{-2}\dot{P}_{\text{NS}}$,
with the period derivative $\dot P_{\text{NS}}$, and the distance to the spin axis $r_{\perp}$. 
Here $\theta$ is the angle between the barycentre of the plate and the spin axis of the NS, and we assume that properties of the vortex line at this position represent that of the whole vortex lines in the plate. 

After the crust of the NS cracking, the plates move and collide with each other as shown in Figure \ref{plate model}.
We assume that the thickness of a plate is equivalent to its size. The volume of a plate can be estimated by $\sim R_{\rm p}^{3}$.
The total collision number per unit time, i.e., the burst rate, can be evaluated by
\begin{align}
\mathcal{R} &\sim N_{\text p} n_{\text p} v_{t} \sigma_{\text p}
\sim \frac{4\pi R_{\text{NS}}^{2} v_t \Delta R}{R_{\rm p}^{4}} 
\simeq 767.7 \,\unit{h^{-1}} \tan \theta 
 \notag \\
&\times \left(\frac{\Delta \Omega/\Omega}{10^{-2}}\right)^2
\left(\frac{\dot P_{\text{NS}}}{10^{-11}\,\unit{s\,s^{-1}}}\right) \left(\frac{\varepsilon_{\rm m}}{10^{-5}}\right)^{-2} 
\left(\frac{P_{\text{NS}}}{10^{-2}\,\unit{s}}\right)^{-5}, 
\label{rate}
\end{align}
with the total number of plates in the crust of the NS 
~\footnote{If we consider the volume of each plate is $V' \sim R_{\rm p}^2 \Delta R$,  then the total number of plates in the crust of the NS $N_{\text p}'  \sim 4 \pi R_{\text{NS}}^{2}\Delta R/(R_{\rm p}^2 \Delta R)$, the number density of plates $n_{\text p}' \sim 1/(R_{\rm p}^2 \Delta R)$, the collision section of plates $\sigma_{\text p}' \sim  R_{\rm p}\Delta R$. Therefore, the corresponding burst rate, $\mathcal{R}' \sim N_{\rm p}' n_{\rm p}' v_{t} \sigma_{\rm p}'\sim 4 \pi R_{\rm NS}^{2} v_{t}/(R_{\rm p}^{3}).$
According to Eq.(\ref{rp}) and Eq.(\ref{rate}), 
$\mathcal{R}'/\mathcal{R} = R_{\rm p}/\Delta R \sim 0.1 ,$
i.e., the burst rate will be one order of magnitude smaller. In reality, there should also be a distribution of thickness of plates, and here we take $R_{\rm p}$ as a typical value.}
$N_{\text p}  \sim 4 \pi R_{\text{NS}}^{2}\Delta R/R_{\rm p}^{3}$, the number density of plates $n_{\text p} \sim 1/R_{\rm p}^{3}$, the collision section of plates $\sigma_{\text p} \sim  R_{\rm p}^2$, the thickness of the crust $\Delta R = 10^5\,\unit{cm}$. 
In the scenario of the magnetic dipole radiation, $\dot P_{\text{NS}} \simeq (8 \pi^{2} R_{\text{NS}}^{6}/3 c^{3} I)B_{\text{NS}}^2P_{\text{NS}}^{-1}$, where $I$ and $B_{\text{NS}}$ are the moment of inertia and the magnetic field of the NS, respectively. 
In general, FRBs may be beamed with a beaming factor $f_{\rm b}$. If the FRBs are
emitted isotropically, the observed event rate of FRBs
would be $\mathcal{R}_{\rm{obs}} \simeq f_{\rm b} \mathcal{R}$ \citep[e.g.,][]{zhang2020binary, yang2021}.
Therefore, Eq.(\ref{rate}) can be represented by
\begin{align}
\mathcal{R}_{\rm{obs}} \sim 749.7 \,\unit{h^{-1}} \tan \theta f_{\rm b} \left(\frac{\Delta \Omega/\Omega}{10^{-2}}\right)^2
\left(\frac{B_{\text{NS}}}{10^{13}\,\unit{G}}\right)^2 \left(\frac{\varepsilon_{\rm m}}{10^{-5}}\right)^{-2}
\left(\frac{P_{\text{NS}}}{0.01\,\unit{s}}\right)^{-6}.
\label{dipole}
\end{align}

\begin{figure*}
\centering
\subfigure[$B_{\rm{NS}} = 10^{13}\,\unit{G}$]{
\label{Fig.sub.1}
\includegraphics[width=0.38\textwidth]{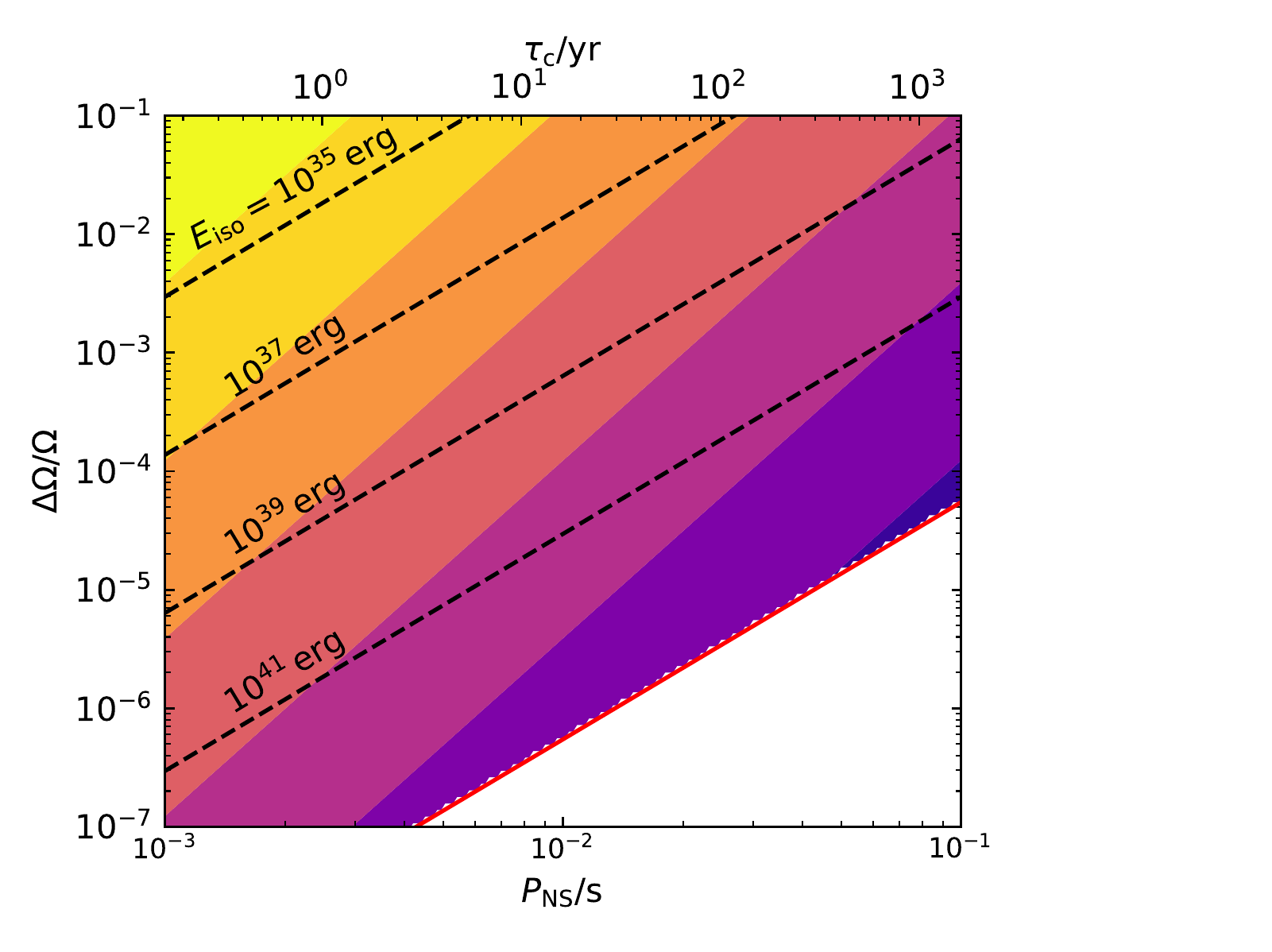}
}\hspace{-17mm}
\subfigure[$B_{\rm{NS}} = 10^{13.5}\,\unit{G}$]{
\label{Fig.sub.2}
\includegraphics[width=0.38\textwidth]{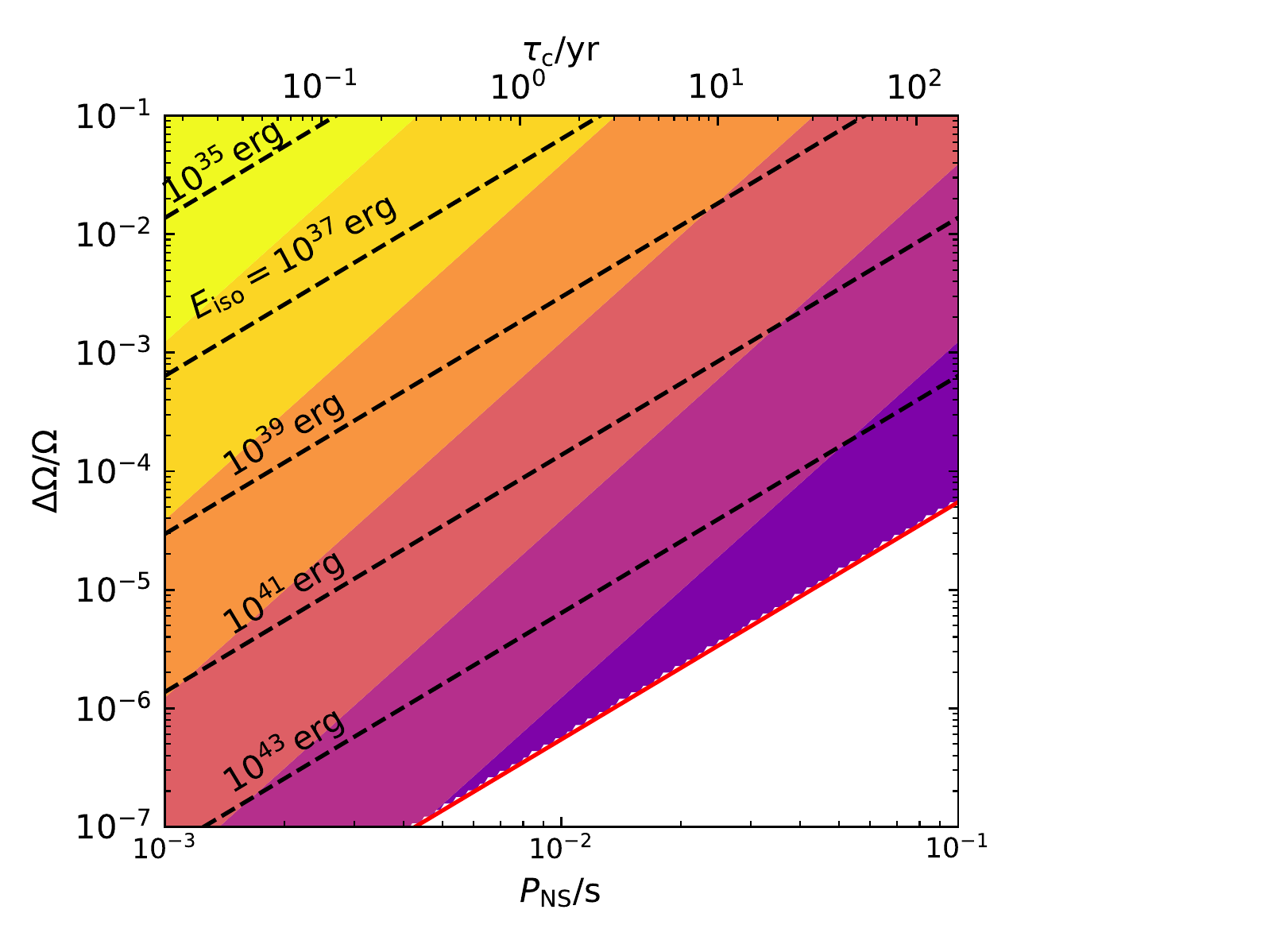}
}\hspace{-17mm}
\subfigure[$B_{\rm{NS}} = 10^{14}\,\unit{G}$]{
\label{Fig.sub.3}
\includegraphics[width=0.38\textwidth]{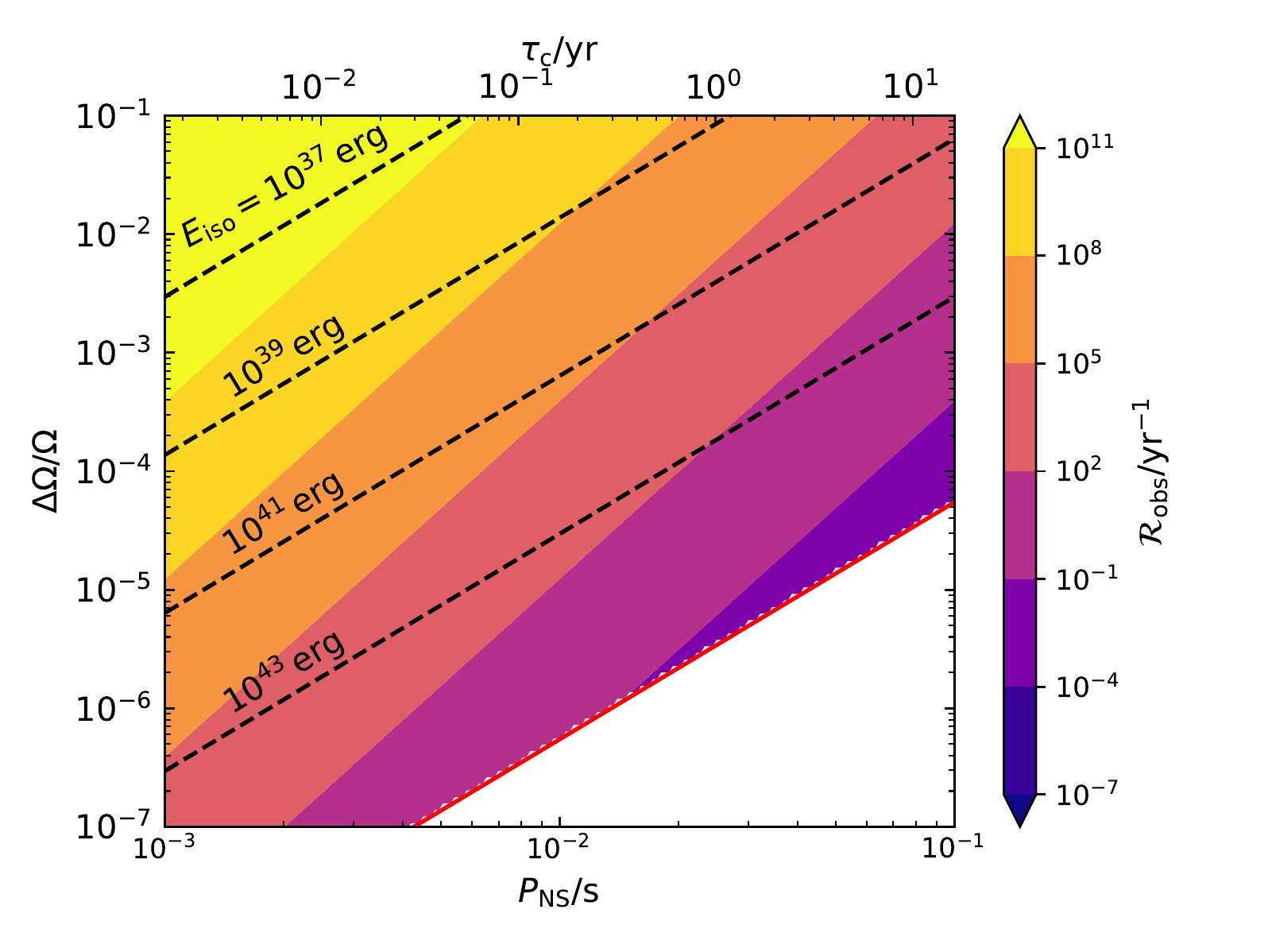}}
\caption{The observed burst rate as a function of period and glitch amplitude of the NS.
Three panels show NSs with different magnetic fields.
Colored regions represent different observed burst rates according to Eq.(\ref{dipole}).
All panels use the same colorscale shown by the colorbar.
Black dashed lines indicate the isotropic energy of FRBs according to Eq.(\ref{Energyiso}) with $\zeta = 1$ and $f_{\rm{b}} = 0.1$.
Red solid lines denote the size of a plate is equal to the radius of the NS $R_{\rm {p}} = R_{\rm{NS}}$, i.e., colored regions satisfy $R_{\rm {p}} < R_{\rm{NS}}$.
The characteristic age is evaluated by $\tau_{\rm {c}} \equiv P_{\text{NS}}/(2\dot P_{\text{NS}})$, with $\dot P_{\text{NS}} \simeq (8 \pi^{2} R_{\text{NS}}^{6}/3 c^{3} I)B_{\text{NS}}^2P_{\text{NS}}^{-1}$.
The other model parameters are taken as $\theta = \pi/4$,  and $\varepsilon_{\rm m} = 10^{-5}$.
}
\label{amplitude}
\end{figure*}

The observed burst rate as a function of period and glitch amplitude of the NS is shown in Figure \ref{amplitude}. Three panels show NSs with different magnetic fields of $10^{13}$, $10^{13.5}$, $10^{14}\,\unit{G}$, respectively.
Different colors denote different observed burst rate $\mathcal{R}_{\rm{obs}}$ with $f_{\rm{b}} = 0.1$. 
The colored regions satisfy that the size of a plate is smaller than the radius of the NS, i.e., $R_{\rm {p}} < R_{\rm{NS}}$.
Therefore, when we get the burst rate of a repeating FRB, we can limit $P_{\text{NS}}$ and $\Delta \Omega/\Omega$ of the central engine. 
On the other hand, we can also infer the possible FRB rate produced by an NS with a certain glitch amplitude, period and magnetic field.

For FRB 121102, the peak burst rate of is $122\,\unit{h^{-1}}$ \citep{lidi21}, i.e., the waiting time of two bursts of about $30\,\unit{s}$. 
According to Eq.(\ref{rate}), the waiting time between two collisions is $t_{\rm w}\simeq 1/\mathcal{R_{\rm obs}} \simeq 48.0\,\unit{s}$, with $f_{\rm b} = 0.1$, $\tan \theta = 1$, $\Delta \Omega/\Omega = 10^{-2}$, $\dot P_{\text{NS}} = 10^{-11}\,\unit{s\,s^{-1}}$, $\varepsilon_{\rm m} = 10^{-5}$, and $P_{\text{NS}} = 0.01\,\unit{s}$.
If taking $\Delta \Omega/\Omega = 10^{-4}$, with the same other parameters, the size of plates is $R_{\text{p}} \sim 7.4 \times 10^4\,\unit{cm}$ and the waiting time is  $t_{\text w} \simeq 5.6 \,\unit{day}$, which is not enough to explain the high burst rate like FRB 121102. 

The interglitch time can be simply estimated by the time to reach the relative rotation velocity when the vortex lines unpinning again after a glitch \citep{pizz2011, haskell2015}
\begin{align}
\tau_{g} \sim \frac{\omega_{c}}{\dot{\Omega}}
\simeq \frac{155.0}{\sin \theta} \,\unit{days}\left(\frac{P_{\text{NS}}}{10^{-2}\,\unit{s}}\right)^3 \left(\frac{B_{\text{NS}}}{10^{13}\,\unit{G}}\right)^{-2}.
\label{glitchtime}
\end{align}

When the plates of the NS collide with each other, this will excite shear mode oscillations. 
At about a few tens of NS radius, there is a charge starved region and FRBs can be produced by coherent radiations \citep{kumar2020, lu2020, yang2020, yang2021}. 
The predicted isotropic energy of FRBs can be evaluated by
\begin{align}
E_{\rm {iso}} \sim \frac{\zeta}{f_{\rm b}} \frac{B_{\text{NS}}^{2}}{8 \pi} R_{\rm p}^{3} & \simeq  1.6 \times 10^{37}\,\unit{erg} \,\zeta \left(\frac{f_{\rm b}}{0.1}\right)^{-1}
\left(\frac{B_{\text{NS}}}{10^{13}\,\unit{G}}\right)^2 \notag \\
& \times \left(\frac{P_{\rm NS}}{10^{-2}\,\unit{s}}\right)^3 \left(\frac{\varepsilon_{\rm m }}{10^{-5}}\right)^{3/2}
\left(\frac{\Delta \Omega/\Omega}{10^{-2}}\right)^{-3/2},
\label{Energyiso}
\end{align}
where $\zeta$ is the magnetic energy conversion fraction.
The pulse width of an FRB triggered by the collision of two plates can be estimated by the propagation time delay of  shear wave between different paths \citep[e.g.,][]{lu2020}
\begin{align}
\Delta t \sim \frac{R_{\text{p}}}{\sqrt{\mu/\rho_{\rm n}}} \simeq 0.2\,\unit{ms} \left(\frac{R_{\rm p}}{10^4\,\unit{cm}}\right).
\label{dt}
\end{align}

\begin{figure}
\centering
\includegraphics[angle=0,scale=0.51]{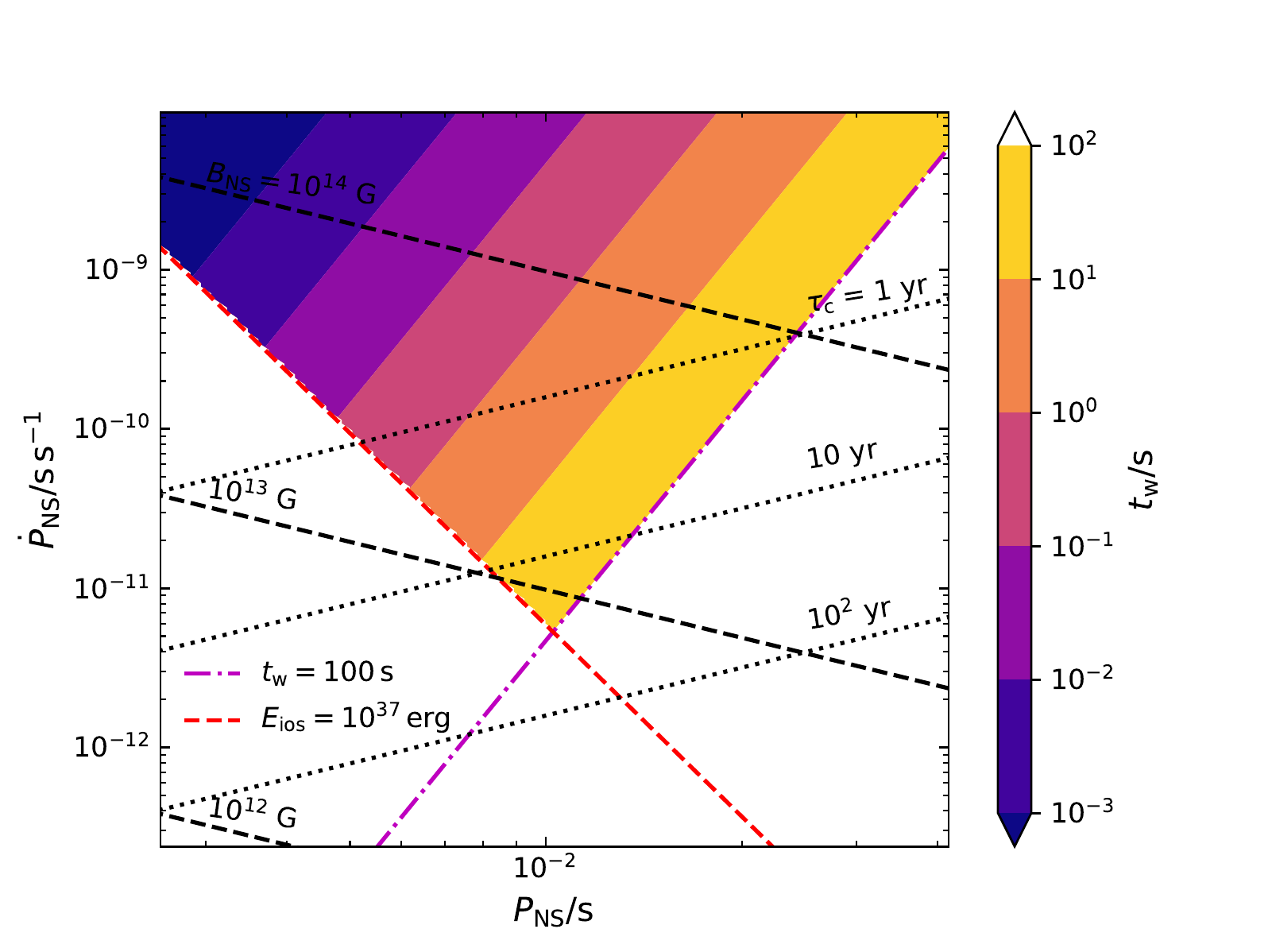}
\caption{Limits on the FRB 121102 source in the $P-\dot P$ diagram.
The red dashed line and purple dash-dotted line denote the waiting time between two bursts $t_{\text{w}} = 100\,\unit{s}$ and the isotropic energy of FRBs $E_{\text{iso}} = 10^{37}\,\unit{erg}$, respectively.
Black dashed lines indicate the magnetic field of the NS $B_{\rm{NS}} = 3.2 \times 10^{19}\,\unit{G}\,(P_{\text{NS}}\dot{P}_{\text{NS}})^{1/2}$, and black dotted lines denote the characteristic age $\tau_{\rm {c}} \equiv P_{\text{NS}}/(2\dot P_{\text{NS}})$.
The enclosed area is the range that meets $t_{\text{w}} < 100\,\unit{s}$, and $E_{\text{iso}} > 10^{37}\,\unit{erg}$, which is suitable to explain the burst rate of FRB 121102.
Different colors denote different $t_{\rm w} \simeq 1/\mathcal{R_{\rm{obs}}}$.
The other model parameters are taken as $\theta = \pi/4$, $\Delta \Omega/\Omega = 10^{-2}$, $\varepsilon_{\rm m} = 10^{-5}$, and $f_{\rm{b}} = 0.1$.}
\label{P-Pdot}
\end{figure}

As shown in Figure 1 and Figure 3 of \cite{lidi21}, the isotropic energy of FRB 121102 is in the range $10^{36.5}-10^{40}\,\unit{erg}$, and the waiting time between bursts is about $100\,\unit{s}$. 
Therefore, for FRBs like FRB 121102 with high burst rate and high enough energy, the limited parameter space of period and period derivative of an NS is shown in Figure~\ref{P-Pdot}.
The area enclosed by the purple dash-dotted line, and red dashed line satisfies the waiting time between two bursts $t_{\text{w}} < 100\,\unit{s}$, and the isotropic energy of FRBs $E_{\text{iso}} > 10^{37}\,\unit{erg}$.
In this area, different colors denote different waiting time of bursts $t_{\text{w}}$. 
Black dashed lines and black dotted lines indicate the magnetic field of the magnetic dipole radiation and the characteristic age, respectively.
The other parameters are taken as $\theta = \pi/4$, $\Delta \Omega/\Omega = 10^{-2}$, $\varepsilon_{\rm m} = 10^{-5}$, and $f_{\rm{b}} = 0.1$.

As shown in Figure~\ref{P-Pdot}, in order to explain the high burst rate like FRB 121102, a magnetar with a magnetic field $B_{\rm NS} > 10^{13}\,\rm{G}$, a spin period $P_{\text{NS}} \sim 10^{-2}\,\rm{s}$, and an age of several decades is required. 
The active time scales of magnetars are about few thousand years \citep{kaspi2017}. Therefore, the fraction of the sources that can produce FRBs like FRB 121102 is $\sim 1\%$. At present, more than 600 FRBs have been observed \citep[e.g.,][]{frbcatnew2020, chimecat2021}, and the burst rate of two FRBs, i.e., FRB 121102 and FRB 20201124A, can be very high. Therefore, such a model might be consistent with the current observations.

Due to a possible young environment, the radiation of the FRBs can be affected by the left-over material with a large free-free opacity after the supernova explosion \citep[e.g.,][]{luan2014, murase2016, metzger2017, yang2019snr}. When the burst propagates through the ejecta shell, the free-free optical depth can be represented by
\begin{align}
\tau_{\mathrm{ff}}= \alpha_{\mathrm{ff}} L_{\rm shell} \simeq\left(0.018 T_{\rm {shell}}^{-3 / 2} Z^{2} n_{e} n_{i} \nu^{-2} \bar{g}_{\mathrm{ff}}\right) L_{\rm shell} ,
\end{align}
where $\alpha_{\mathrm{ff}}$ is the free–free absorption coefficient, $L_{\rm shell}$ is the supernova remnant (SNR) thickness, $T_{\rm shell}$ is the temperature of the ejecta shell, $Z \sim 1$ is the charge of ions, $n_{e}$ is the electron density, $n_{i}$ is the ion density in the SNR, $\nu$ is the is the frequency, $\bar{g}_{\mathrm{ff}} \sim 1$ is the Gaunt factor for the free–free emission, and $L_{\rm shell}$ is the SNR thickness.  
Electron density $n_e$ can be calculated by $n_{e} \simeq M_{\rm SNR} /(4 \pi \mu_{m} m_{p} r_{\rm SNR}^{2} L_{\rm shell})$, with the mean molecular weight $\mu_m = 1.2$ for a solar composition in the ejecta, where $m_p$ is the proton mass, $r_{\rm SNR} \sim v_{\rm SNR}t_{\rm SNR}$ is the SNR radius, $v_{\rm SNR} = \sqrt{2E_{\rm SNR}/M_{\rm {SNR}}}$ is the SNR ejecta velocity, $E_{\rm SNR}$ is the SNR kinetic energy, and $t _{\rm SNR}$is the SNR age. 
Taking $n_e \simeq n_i$ and $L_{\rm shell} \sim r_{\rm SNR} $, the free-free optical depth can be estimated by
\begin{align}
\tau_{\mathrm{ff}} & \simeq  3600\left(\frac{T_{\rm {shell}}}{10^{4} \mathrm{~K}}\right)^{-3 / 2}\left(\frac{M_{\rm {SNR}}}{M_{\odot}}\right)^{9 / 2} \\ \notag
& \times \left(\frac{E_{\rm {SNR}}}{10^{51} \mathrm{erg}}\right)^{-5 / 2} \left(\frac{t_{\rm SNR}}{1 \mathrm{yr}}\right)^{-5}\left(\frac{\nu}{1 \mathrm{GHz}}\right)^{-2}.
\end{align}
When $\tau_{\mathrm{ff}} < 1$, the ejecta with $T_{\rm shell} \sim 10^4 \, \rm {K}$ will become optically thin for FRBs at $\nu \sim 1\,\rm{GHz}$ after 
\begin{align}
t_{\rm SNR} \gtrsim 5 \mathrm{yr}\left(\frac{M_{\rm SNR}}{M_{\odot}}\right)^{9 / 10}\left(\frac{E_{\rm SNR}}{10^{51} \mathrm{erg}}\right)^{-1 / 2}.
\end{align}
Therefore, FRBs can be observed few years later after the SN explosion. 
Furthermore, the interaction of the supernova ejecta or the compact binary mergers ejecta with the circumstellar medium can explain the evolution of FRB 121102 dispersion measure of ${\rm{d\,DM}}/{\rm{d}}t \sim 0.85\pm 0.1\,\rm{pc\,cm^{-3}yr^{-1}}$ \citep[e.g.,][]{lidi21, zhao2021}, the rotation measure of dropping by an average of $15\%\,\rm{yr^{-1}}$ \citep[e.g.,][]{piro2018, hilmarsson2021_121102, zhao2021}, and the estimated age is several decades \citep[e.g.,][]{hilmarsson2021_121102, zhao2021}. 

\subsection{Waiting Time and Energy Distributions of Repeating FRBs}\label{sec2.3}

Next, we consider that the sizes of plates have a dynamic devolution.
When the crust of an NS cracks gradually from large to small pieces, the mean size of the plates would decrease gradually, leading to an increasing burst rate according to Eq.(\ref{rate}). Conversely, when the plates bond together, the mean size would increase, leading to a relatively small burst rate.
In this section, we derive the possible distribution of the burst rate considering the process of plate fragmentation.
When the crust cracks, the breaking time scale $T$ to make a plate turn into small size is of order of the magnitude of the timescale of the shear wave propagating in the plate.  
Because the shear wave speed can be approximate to a fixed value, the breaking time scale can be estimated by $T \propto R_{\text{p}}$. 
To simplify, we consider that a large plate breaks into two small plates with the same size each time.
The initial size of a plate is $R_{\text{p}, 0}$. When the plate size become $R_{\text{p}, n}$ after the $n$-th time crack, it will take a time of
\begin{align}
t = \sum_{i = 0}^{n-1} T_{i}= \sum_{i = 0}^{n-1}(1/2^i) T_0 = 2T_0(1 - 1/2^n),
\label{platetime}
\end{align}
where $T_i$ is the time to make the plates into two smaller plates, and $T_0$ is the initial breaking time of a plate size $R_{\text{p}, 0}$.
At the same time, the size of plates after the $n$-th time crack satisfies 
\begin{align}
R_{\text{p}, n} = (1/2^n)R_{\text{p}, 0}.\label{platesize}
\end{align}
Therefore, according to Eq.(\ref{platetime}) and Eq.(\ref{platesize}) the relation between time and the burst rate satisfies 
\begin{align}
t \propto (1 - R_{{\text p}, n}/R_{\text{p}, 0}) \propto (1 - A\mathcal{R}^{-1/4}),
\end{align}
with a constant $A$.
We define the distribution of the burst rate as $f(\mathcal{R})$ and the distribution of the sampling time that calculates the burst rate as $f(t)$.
Assuming that the sampling time is uniform random, i.e., $f(t) = \text{const}$, the distribution of the burst rate can be estimated as
\begin{align}
f(\mathcal{R}) = f(t)\text{d}t/\text{d}\mathcal{R}\propto \mathcal{R}^{-5/4}.
\label{fratemodel}
\end{align}

After a certain amount of time of the plate cracking, due to the fusion process of the plates, some small plates would bond into large plates.
The typical timescale of the plate fusion might satisfy $T \propto R_{\text{p}}$.
Therefore, for the fusion process as the inverse process of the plate cracking, the distribution of the burst rate can also be represented by Eq.(\ref{fratemodel}).

When the size of cracking plates reaches a critical value, the plates will not further crack due to the balance of the fragmentation and fusion processes. 
Therefore, there might be a cutoff of the burst rate.
A general form of the distribution of the burst rate might be represented by 
\begin{align}
f(\mathcal{R}) = C_{\mathcal{R}} \mathcal{R}^{\alpha_{\mathcal{R}}} \exp \left(-\frac{\mathcal{R}}{\mathcal{R}_{0}}\right),
\label{frate}
\end{align}
where $\alpha_{\mathcal{R}}$ is the power-law index, $C_{\mathcal{R}}$ is the normalization constant, i.e., $\int C_{\mathcal{R}} \mathcal{R}^{\alpha_{\mathcal{R}}} \exp \left(-\mathcal{R}/\mathcal{R}_{0}\right) \text{d} \mathcal{R} = 1$, and $\mathcal{R}_0 = \int_{0}^{\infty} \mathcal{R} f(\mathcal{R}) \text{d} \mathcal{R}$ is the mean burst rate as the cutoff position.

If the burst rate is constant, the waiting time distribution satisfies the Poisson interval distribution~\footnote{The distribution of waiting time can also be described by the Weibull distribution, with the shape parameter $k$ \citep[e.g.,][]{zhangguoqiang2021}. Bursts show clustering behavior when $k < 1$, and if $k = 1$, the Weibull distribution reduces to the Poisson distribution.
\cite{oppermann2018} derived $k \simeq 0.34$ using 17 bursts from the Robert C. Byrd Green Bank Telescope and Arecibo telescopes.
With the largest number of samples observed by the same telescope, \cite{zhangguoqiang2021} found $k \simeq 0.72$ for $t_{\text{w}} > 30\,\unit{ms}$.
However, the waiting time might be affected by many factors, such as the telescope sensitivity, the radiation properties of FRBs.
Here, we assume the waiting time satisfies a Poisson distribution for a general discussion.}
\citep{wheatland1998} $P(t_{\text w}) = \mathcal{R}e^{-\mathcal{R}t_{\text w}}$.
When the burst rate is time-dependent, in discrete time intervals, $\mathcal{R}$ can also be considered as constant \citep{zhangguoqiang2021}, and the distribution of the waiting time can be derived by \citep{aschwanden2011}
\begin{align}
P(t_{\text w}) = C_{t_{\text w}} \frac{\int_{0}^{\infty} f(\mathcal{R}) \mathcal{R}^{2} e^{-\mathcal{R} t_{\text w}} \text{d} \mathcal{R}}{\int_{0}^{\infty} \mathcal{R} f(\mathcal{R}) \text{d} \mathcal{R}},
\label{ptw}
\end{align}
where $C_{t_{\text w}}$ is the normalization constant.
According to Eq.(\ref{frate}), the distribution of the waiting time can be represented by
\begin{equation}
P(t_{\text w}) = C_{t_{\text w}} \frac{\mathcal{R}_{0}^{\alpha_{\mathcal{R}} + 2} }{\left(1 + \mathcal{R}_{0} t_{\text w}\right)^{\alpha_{\mathcal{R}} + 3}} \Gamma(\alpha_{\mathcal{R}} + 3),
\label{ptwalpha}
\end{equation}
where $\Gamma(x)$ is the Gamma function. 
In the plate collision scenario, the distribution of the waiting time according to
Eq.(\ref{fratemodel}) and Eq.(\ref{ptwalpha}) can be written as
\begin{equation}
P(t_{\text w}) = C_{t_{\text w}} \frac{\mathcal{R}_{0}^{3/4} }{\left(1 + \mathcal{R}_{0} t_{\text w}\right)^{7/4}} \Gamma\left(7/4 \right).
\end{equation}

\begin{figure}
\includegraphics[width=\columnwidth]{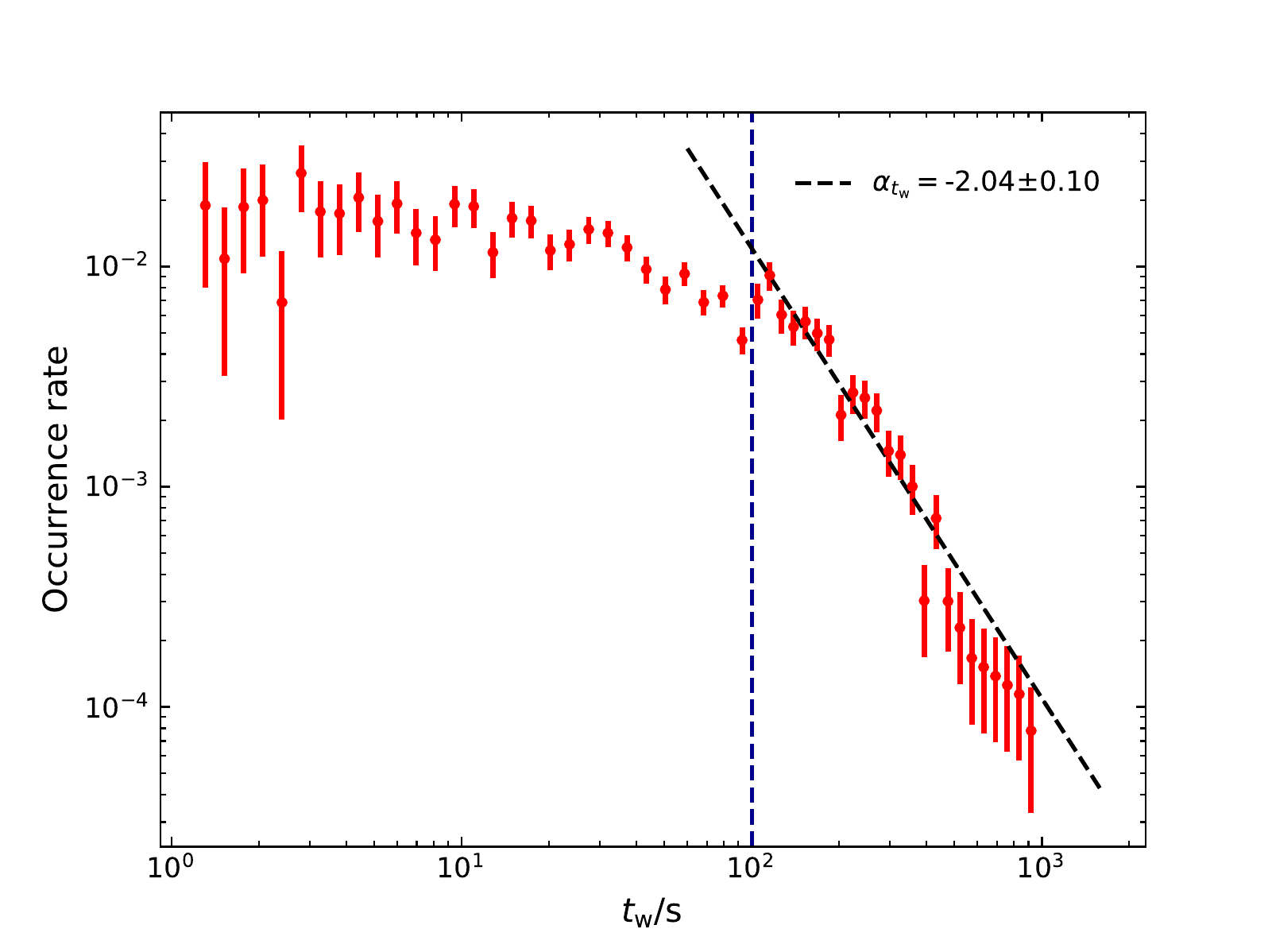}
\caption{
The waiting time occurrence rate for FRB 121102.
The data is from \protect \cite{lidi21} and the occurrence rate is shown as red points.
The best-fitting results are shown as the black-dashed lines with the power-law index $\alpha_{t_{\text{w}}} = -2.04 \pm 0.10$. The blue dashed vertical line is the breakpoint of waiting time of about $100\,\unit{s}$, corresponding to the mean rate of FRB 121102.}
\label{121102waitingtime}
\end{figure}

There is a large uncertainty in the statistics of the burst rate for different time intervals. For a large time interval, the information of short-term burst rate will be erased.
Therefore, we focus on making statistics on the waiting time.
For large waiting time, Eq.(\ref{ptwalpha}) can approach a power-law form, i.e., $P(t_{\text w}) \propto t_{\text w}^{\alpha_{t_{\text{w}}}}$, with $\alpha_{t_{\text{w}}} = -\alpha_{\mathcal{R}} - 3$. 
The waiting time distribution of FRB 121102 has two components. One is at a few milliseconds, which might be caused by substructures of some bursts \citep{lidi21}.
Here we only consider the occurrence rate of the waiting time from $1\,\unit{s}$ to $10^3\,\unit{s}$ of FRB 121102 \citep{lidi21}, shown in Figure~\ref{121102waitingtime}.
We can see that the breakpoint of FRB 121102 is about $t_{\text{w}, 0} = 1/\mathcal{R}_0 \sim 100\,\unit{s}$. 
And the best-fitting power-law index is $\alpha_{t_{\text{w}}} = -2.04 \pm 0.10$, using the Markov chain Monte Carlo (MCMC) method \citep{foreman2013}.
Therefore, the difference of the power-law index $\alpha_{t_{\text w}}$ in the plate collision scenario and observations is about $16.6\%$.

Furthermore, we consider that the plate sizes have a power-law distribution, and calculate the burst energy distribution.
When the plates crack, the characteristic size can be estimated by Eq.(\ref{rp}). 
Therefore, the burst energy in Eq.(\ref{Energyiso}) induced by collisions between plates also can be represented by a power-law distribution. 
In order to get the energy distribution of bursts, we assume that the size of the plates has a power-law distribution. 
The number of plates in the range between $R_{\text p}$ and $R_{\text p} + \text{d}R_{\text p}$ can be represented by $N(R_{\text p})\text{d}R_{\text p}\propto R_{\text p}^{\alpha_{R}}\text{d}R_{\text p}$. 

In astrophysics systems, many phenomena can be represented by a power-law function, such as the energy distribution of solar flares, X-ray bursts of magnetars, X-ray flares of gamma-ray bursts, giant pulses of Crab pulsar \citep[e.g.,][]{gogus2000, popov2007, wang2013, wangfayin2015, chengyingjie2020, lyufen2021, yangyh2021}.
The power-law distributions are a natural predication of the self-organized criticality (SOC), which was first introduced by \cite{bak1987, bak1988}.
The basic idea of SOC can be illustrated with a pile of sand. Suppose that we build up a pile by adding sand with a grain at a time. The slope of the pile will evolve to a critical angle, at which the pile collapses and avalanches occur. The distribution of the size of avalanches can be represented by a power law \citep{bak1988}.
In Euclidean dimension $S = 3$, the avalanche size $L$ distribution can be expressed by $N(L)\text{d}L\propto L^{-3}\text{d}L$ \citep{asc2012}. 
Therefore, we assume the power-law index of the size of plates $\alpha_{R} = -3$.
The energy distribution of bursts $N(E) \text{d}E \propto E^{\alpha_{E}}\text{d}E$ using the relationship $E \propto R_{\text p}^3$ (Eq.(\ref{Energyiso})) can be represented by
\begin{equation}
N(E)\text{d}E =N(R_{\text p}(E))\frac{\text{d}R_{\rm p}}{\text{d}E}\text{d}E \propto E^{-5/3}\text{d}E.
\label{N(E)}
\end{equation}

The statistical results, such as the energy distribution, can be used to test the proposed models and deepen our understanding of the possible mechanism to produce FRBs.
Recently, the energy distribution of FRB 121102 has been widely studied.
\cite{wangfayin2017} found that $\alpha_{E} = -1.8 \pm {0.15}$ for FRB 121102. 
\cite{wangweiyang2018} studied that $\alpha_{E} = -2.16 \pm {0.24}$. With a larger sample by different radio telescopes at different frequencies, \cite{wangfayin2019} shown that $\alpha_{E}$ is in the range form $-1.8$ to $-1.6$.
Recently, \cite{lidi21} reported the largest number of FRB 121102 samples observed by the same telescope and $\alpha_{E}$ at $E > 10^{38}\,\unit{erg}$ can be represented by $-1.86\pm 0.02$ \citep{zhangguoqiang2021}.
In the plate collision model, the power-law index of the energy distribution of bursts is $\alpha_{\text E} \simeq -1.67$ according to Eq.(\ref{N(E)}), which is consistent with observations of FRB 121102.

\section{Discussion and Conclusions}\label{sec3}

Recently, more and more high-burst-rate repeating FRBs have been detected, in which the most outstanding one is FRB 121102 with the average waiting time of about $100\,\unit{s}$ and a peak burst rate more than $122\,\unit{h^{-1}}$ \citep{lidi21, zhangguoqiang2021, jahns2022}.
In this work, we propose that the plate collision model can be used to explain repeating FRBs with high burst rates. 
Due to the spin evolution of the NS, the relative rotation velocity between the superfluid neutrons and the NS lattices will increase, leading to the stress increasing in the pinning region of the crust. 
Then the crust may crack into plates when the stress reaches a critical value, each of which moves and collides with its adjacent ones leading to FRBs triggered.
We focus on the discussion of the influence of the NS spin evolution and the vortex pinning-induced crust cracking.
In this scenario, for the observed high burst rate of $\sim100~{\rm h^{-1}}$ for FRB 121102, one requires an NS with the spin period of $P_{\text{NS}} \sim 0.01\,\unit{s}$, the period derivative of $\dot P_{\text{NS}} \sim 10^{-11}\,\unit{s\,s^{-1}}$, the glitch amplitude $\Delta \Omega/\Omega \sim 10^{-2}$, and the breaking strain $\varepsilon_{\text{m}} \sim 10^{-5}$. 
For the most active glitching pulsar, PSR J0537-6910, with $P_{\text{NS}} \approx 16\,\unit{ms}$, $\dot P_{\text{NS}} \approx 5.2 \times 10^{-14}\,\unit{s\,s^{-1}}$, and $\Delta \Omega/\Omega \simeq 10^{-7}$ \citep[e.g.,][]{marshall1998, marshall2004, espinoza11}, in the plate collision scenario, the possible plate size is $2.3 \times 10^{6}\,\unit{cm}$, which is larger than the radius of the NS. It means that there might be very low probability for PSR J0537-6910 to produce FRBs in our scenario.

In the plate collision scenario, a very young NS, $\sim 10 - 100\,\unit{yr}$, is needed to explain the high burst rate like FRB 121102. 
In the studies of NS cooling \citep[e.g.,][]{lattimer1994}, for a normal NS $\sim 1 - 10\,\rm{yr}$ after its birth, the crust temperature $T_{\text{crust}} \sim  10^9\,\rm{K}$; at the age of $\sim 10\,\rm{yr}$, the crust temperature will have a rapid cooling due to the neutrino emission; after few decades cooling, the crust temperature reaches $T_{\text{crust}} \sim  10^7 - 10^{8}\,\rm{K}$.
In the study of the temperature distribution in magnetized NS crusts \citep{geppert2004}, they find the crustal temperature difference, at the density of $\sim 10^{11} - 10^{14}\,\rm{g\,cm^{-3}}$, is an order of magnitude.
If it is considered that when the crust temperature is lower than $0.1T_{\text{m}}$, where $T_{\text{m}}$ is the crystal melting temperature, the crust can change from plastic flow to brittle response \citep{ruderman2}. The crystal melting temperature $T_{\text{m}} \sim 1.4 \times 10^{10} \rho_{14}^{1/3} \,\rm{K}$, with $\rho_{14} = 10^{14} \mathrm{~g} \mathrm{~cm}^{-3} $ \citep[e.g.,][]{blaes1989, chamel2008}, i.e., $T_{\rm{m}} \sim 10^{9} - 10^{10}\,\rm{K}$ for the crust of NSs with a density of $\sim 10^{11} - 10^{14}\,\rm{g\,cm^{-3}}$. Therefore, the evolution of the crustal temperature only plays an important role for extremely young NS with ages less than $\sim 10\,\rm{yr}$ \citep[e.g.,][]{lattimer1994}, which does not significantly impact out model.

The spin evolution of an NS is caused not only by its magnetic dipole radiation, but also by some external forces, such as the accretion process.
In the binary system, there is an accretion torque $N_{\rm acc}$ on the NS due to the interaction between the accreted material and the NS magnetosphere \citep[e.g.,][]{blondin1990, frank2002}.
If $N_{\rm acc}$ is dominant relative to the torque due to magnetic dipole or gravitational-wave radiation of the NS.
The accretion torque is related to the accretion rate. 
The change rate of angular velocity of the NS $\dot \Omega$ can be derived from $N_{\rm acc} = I_0 \dot \Omega$.
$N_{\rm acc}$ can be represented by 
$
N_{\rm acc} \simeq \pm\dot{M}\left(G M_{\text{NS}} R_{\mathrm{in}}\right)^{1 / 2}
$ 
\citep[e.g.,][]{ghosh1979b}, where $\dot{M}$ is the accretion rate of the NS, and $R_{\mathrm{in}} = \xi R_{\text{A}}$ is the interaction radius between the accreted material and the NS magnetosphere, with the Alfv\'en radius $R_{\text{A}} = [\mu_{\text{NS}}^4/(2GM_{\text{NS}}\dot{M}^2)]^{1/7}$ and the factor $\xi \sim 0.01 - 1$ \citep{ghosh1979a, ghosh1979b, filippova2017}.
Then $R_{\text{in}} \simeq 4.1\times10^7 \,\unit{cm} ( \xi/0.1)(B_{\text{NS}}/10^{13}\,\unit{G})^{4/7}  (\dot{M}/\dot{M}_{\text{Edd}})^{-2/7}$, with the rate of Eddington accretion $\dot M_{\text{Edd}} \simeq 10^{18} \,\unit{g\,s^{-1}}$, and the radius and mass of the NS $R_{\text{NS}}/10^6\,\unit{cm}$ and $M_{\text{NS}} = 1.4M_{\odot}$, respectively.
If $R_{\mathrm{in}} < R_{\text{lc}}=c/\Omega  \simeq 4.8 \times 10^7 \,\unit{cm} ( P_{\text{NS}}/0.01\,\unit{s})$, with the radius of the light cylinder $R_{\text{lc}}$, there will exist the interaction between the magnetosphere of the NS and the accreted material around the NS. 
Therefore, in an accretion binary system, the burst rate in Eq.(\ref{rate}) can be estimated by 
$\mathcal{R} \simeq 0.1  \,\unit{h^{-1}}\tan \theta(\dot M/\dot M_{\rm {Edd}})^{6/7}( \xi/0.1)^{1/2}\times(B/10^{13}\,\unit{G})^{2/7}(P/0.01\,\unit{s})^{-3}$, corresponding to the waiting time between two bursts of about $0.4\,\unit{day}$.
The other parameters are taken as $\Delta \Omega/\Omega = 10^{-2}$, $\varepsilon_{\rm m} = 10^{-5}$.
In a binary system, even if the accretion rate reaches $\dot{M}_{\text{Edd}}$, it is not enough to explain the high burst rate of FRB 121102.
Therefore, in our scenario, FRB 121102 with extreme burst rate may not be produced in the binary system.
However, this scenario may explain the repeating sources with a lower burst rate.

On the other hand, we also study the waiting time and energy distributions of repeating FRBs in the scenario of plates cracking and collision. We find that the distribution of burst rate satisfies $f(\mathcal{R}) \propto \mathcal{R}^{-5/4}$.
Assuming that the waiting time distribution $P_{t_{\text{w}}}$ can be estimated by the Poisson process, we find that $P_{t_{\text{w}}} \propto t_{\text{w}}^{-1.75}$ for the large waiting time. And the best fitting power-law index of the waiting time distribution of FRB 121102 is $\alpha_{t_{\text{w}}} = -2.04 \pm 0.10$.
Furthermore, if we assume the size of plates follows a distribution of $N(R_{\text p}) \text{d}R_{\text p}\propto R_{\text p}^{-3}\text{d}R_{\text p}$ according to the SOC theory, we can make an estimation of the burst energy distribution, $N(E) \text{d}E \propto E^{-5/3}\text{d}E$, with $E \propto R_{\text{p}}^3$. 
And many statistical studies on the energy distribution of FRB 121102 find the power-law index is from $-2.16$ to $-1.6$ .
Therefore, the plate collision model is consistent with the observations. 

\section*{Acknowledgements}
We thank Qian-Cheng Liu, Bin Hong, Guo-Qiang Zhang and Zhao-Yang Xia for helpful discussions. 
This work was supported by the National Key Research and Development Program of China (grant No. 2017YFA0402600), the National SKA Program of China (grant No. 2020SKA0120300), the National Natural Science Foundation of China (grant Nos. 11833003, U1831207 and 12003028), and the Project funded by China Postdoctoral Science Foundation (grant No. 2021M703168).

\section*{Data Availability}
This theoretical study did not generate any new data.

\bibliographystyle{mnras}

\bsp	
\label{lastpage}
\end{document}